\documentclass[aps,pre,showpacs,twocolumn]{revtex4}
\usepackage{graphicx,color,epsfig}
\usepackage{amssymb,amsmath}
\usepackage{url}
\usepackage[colorlinks=true,allcolors=blue]{hyperref}

\begin{document}
\title{Chimeras and clusters in networks of hyperbolic chaotic oscillators}
\author{A. V. Cano and M. G. Cosenza}
\affiliation{Grupo de Caos y Sistemas Complejos, Centro de F\'isica Fundamental, Universidad de Los Andes, M\'erida, Venezuela}
\date{Phys. Rev. E \textbf{95}, 030202(R) (2017)}

\begin{abstract}
We show that chimera states, where differentiated subsets of synchronized and desynchronized dynamical elements coexist,
can emerge in networks of hyperbolic chaotic oscillators 
subject to global interactions. As local dynamics we employ Lozi maps which possess hyperbolic chaotic attractors.
We consider a globally coupled system of these maps and 
use two statistical quantities to describe its collective behavior: the average fraction of elements belonging to clusters and
the average standard deviation of state variables. 
Chimera states, clusters, complete synchronization, and incoherence are thus characterized on the space of parameters of the system.
We find that chimera states are related to the formation of clusters in the system. 
In addition, we show that chimera states arise for a sufficiently long range of interactions in nonlocally
coupled networks of these maps. Our results reveal that, under some circumstances, hyperbolicity does not impede the formation of chimera states 
in networks of coupled chaotic systems, as it had been previously hypothesized. 
\end{abstract}

\maketitle

\section{Introduction}
There has been much recent interest in the investigation of the conditions for 
the existence of chimera states (or chimeras) in networks of interacting identical oscillators.  
A chimera state occurs, in general, when the symmetry of the system
of oscillators is broken into coexisting synchronized and desynchronized subsets.
Such states were first recognized in systems of nonlocally coupled phase oscillators \cite{Kuramoto,Abrams}
and have since been the subject of many investigations in a diversity of models,
including coupled map lattices \cite{Omel,Semenov}, chaotic flows \cite{Omel2}, neural systems \cite{Kanas,Omel3}, population
dynamics \cite{Dutta}, van der Pol oscillators \cite{Ulo}, Boolean networks \cite{Rosin}, lasers \cite{Rohm}, and
quantum systems \cite{Bastidas,Vie}.
Chimera states have been also observed experimentally in coupled populations of chemical oscillators
\cite{Showalter,Nkomo}, optical light modulators \cite{Roy}, coupled lasers \cite{Hart}, mechanical \cite{Martens,Kap,Blaha},
electrochemical \cite{Kiss}, and electronic \cite{Larger} oscillator systems. 
Furthermore, chimeras can occur in systems with local (nearest-neighbors) interactions \cite{Clerc,Bera,Hiz} 
or global (all-to-all) interactions \cite{Sen,Pik,Schmidt}.
In fact, Kaneko observed a chimera behavior 
in a globally coupled map network \cite{Kaneko1},
consisting of the coexistence of one synchronized cluster and a cloud of desynchronized elements. This behavior
has been recently identified as a chimera state \cite{Pik,Kaneko2}. Applications of chimera states may arise in real-world phenomena such as 
the unihemispheric sleep in birds and dolphins \cite{Lima}, neuronal bump states \cite{Laing,Sakaguchi},
epileptic seizure \cite{Roth}, power grids \cite{Fila}, and social systems \cite{JC}. 
Reviews of this growing field of research can be found in \cite{Panaggio,Yao,Scholl}.

Although no universal mechanism for their emergence has yet been established, chimera states 
appear in many spatiotemporal dynamical systems under a broad range of conditions, including 
a variety of network topologies and local dynamics.
However, it has been recently argued that chimera states cannot be obtained in networks of oscillators 
possessing hyperbolic chaotic attractors \cite{Semenova,Zakharova}. 
This type of chaotic attractor exhibits a homogeneous structure over a finite range of parameters.
In this Rapid Communication we revisit this hypothesis. 
We consider a network of globally coupled Lozi maps as a prototype of a system possessing hyperbolic chaotic attractors,
and find that chimera states can actually take place for several values of parameters.
These states appear related to the phenomenon of dynamical clustering typical of systems with global interactions. 
To characterize the collective behavior on the space of parameters of the system, we employ two statistical quantities 
that allow us to distinguish between chimera states, clusters, incoherence, and complete synchronization. 
In addition, we show that chimera states can arise for a sufficiently long range of interaction in nonlocally coupled networks of Lozi maps.   

\section{Chaotic hyperbolic maps}
Hyperbolic chaotic attractors possess the property of robust chaos: 
i.e. there exist a neighborhood in the space of parameters of the system where periodic windows are absent and the chaotic attractor is unique.
It has been found that several dynamical systems can display robust chaos; for a review see Ref.~\cite{Sprott}. 
Robustness is an important feature in applications that require reliable operation in a chaotic
regime, in the sense that the chaotic behavior cannot be destroyed by arbitrarily small
perturbations of the system parameters. For instance, networks of coupled maps with
robust chaos have been efficiently employed in communication schemes \cite{Garcia}.

\begin{figure}[h]
\includegraphics[scale=0.32]{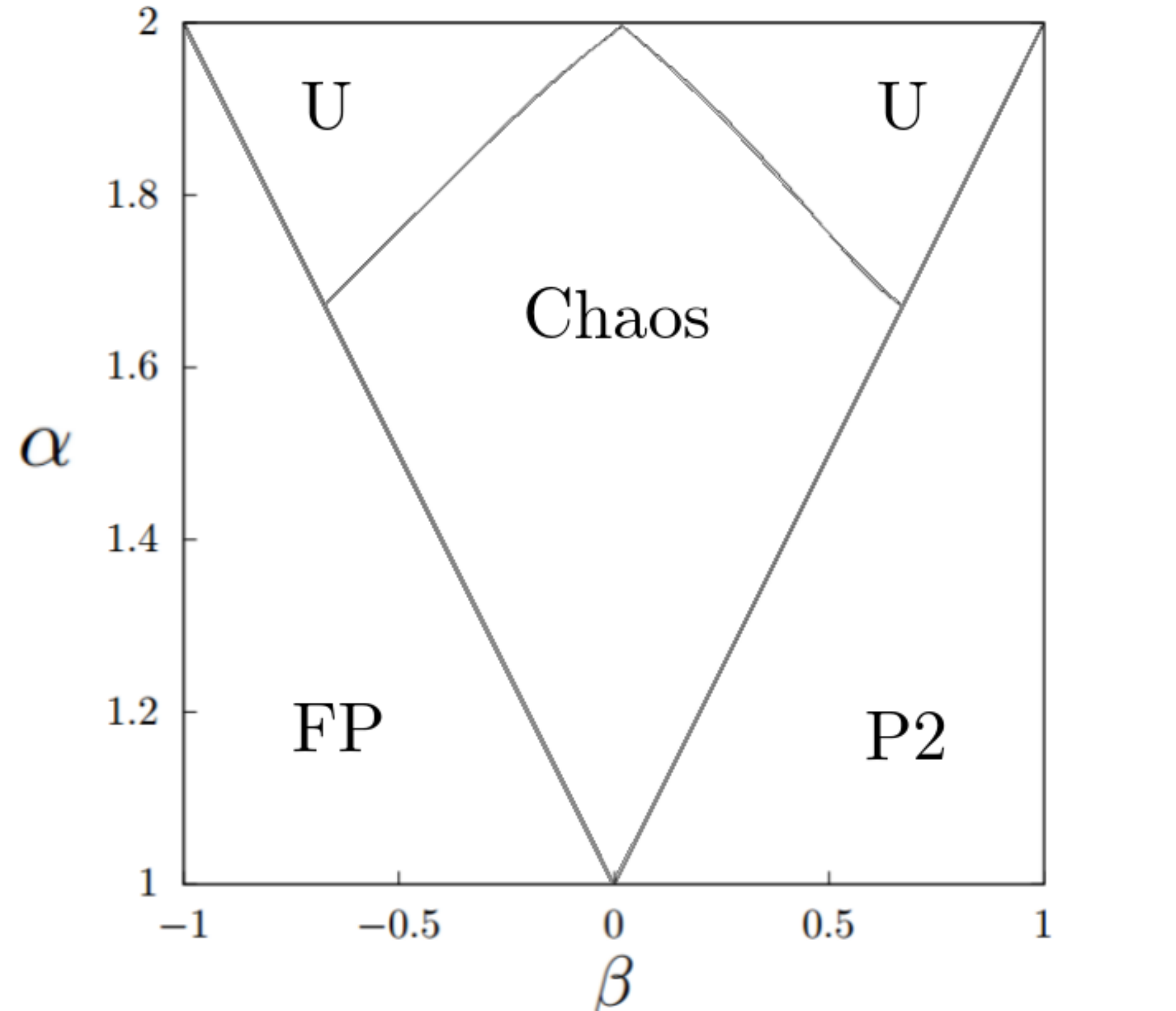}
\caption{Behavior of the Lozi map on the space of parameters $(\alpha,\beta)$. 
Different regions of stable states are indicated: FP, fixed point; 
P2, period-2 orbit; and chaos (robust chaos). The regions marked U correspond to unbounded orbits.}
\label{f1}
\end{figure}

As an example of a hyperbolic chaotic system, we consider the Lozi map \cite{Lozi},
\begin{equation}
 \label{Lozi1}
\begin{array}{lll}
x_{t+1}&=& 1-\alpha \vert x_t \vert +y_t \equiv f(x_t,y_t),  \\
y_{t+1}&=& \beta x_t,
\end{array}
\end{equation}
where $\alpha$ and $\beta$ are real parameters. Figure~(\ref{f1}) shows the
behavior of the Lozi map on the space of parameters $(\alpha,\beta)$.
A stable fixed point exists in the region  $\beta>-1$, $\alpha<1-\beta$, and $\alpha>\beta-1$, while a stable period-2
orbit occurs in the region $0<\beta<1$, $\alpha <1+\beta$, and $\alpha>1-\beta$ \cite{Botella}. 
Robust chaos, characterized by a continuous positive value of the largest Lyapunov exponent of the map~(\ref{Lozi1}), 
takes place on a bounded region of the parameters $\alpha$ and $\beta$, as shown in Fig.~(\ref{f1}). The topology of the chaotic 
attractor is not altered in this region of parameters \cite{Ani}. 

\section{Clusters and chimeras in globally coupled Lozi maps} 
Global interactions in a system occur when all its elements
are subject to a common influence whose origin can be external or endogenous. 
Here we consider the autonomous system of globally coupled Lozi maps described by the equations
\begin{eqnarray}
\label{SystL1}
x_{t+1}^i &=& (1-\epsilon) f(x_t^i,y_t^i) + \epsilon h_t, \\
y_{t+1}^i &=& \beta x_t^i, \\
h_t &\equiv& \frac{1}{N} \sum_{j=1}^N f(x_t^j,y_t^j),  
\label{SystL2}
\end{eqnarray}
where $x_i^t$ and $y_i^t$ give the state variables of map $i$ ($i=1,\ldots,N$) at discrete time $t$, 
the function $f(x_t,y_t)$ is defined in Eq.~(\ref{Lozi1}), and the parameter $\epsilon$ represents the strength 
of the global coupling of the maps. The form of the coupling in Eq.~(\ref{SystL1}) is assumed in the usual diffusive form.

Synchronization in the system of equations~(\ref{SystL1})-(\ref{SystL2}) at time $t$ arises when $(x^i_t,y_t^i)=(x^j_t,y_t^j)$,  $\forall i,j$. 
Note that synchronization of the $x$ variable implies synchronization of the $y$ variable. 
Besides synchronization, the following collective states can be defined in the globally coupled system of equations~(\ref{SystL1})-(\ref{SystL2}).

(i) \textit{Clustering}. A dynamical cluster is defined as a subset of elements that are synchronized among themselves. 
In a clustered state, the elements in the system segregate into $K$ distinct subsets that evolve in time;
i.e., $x^i_t=x^j_t=X^\nu_t$, $\forall i,j$ in  the $\nu$th cluster, with $\nu=1,\ldots,K$. 
We call $n_\nu$ the number of elements belonging to the $\nu$th cluster; then its relative size is $p_\nu=n_\nu/N$. 
 
(ii) \textit{Chimera state}. A chimera state  consists of the coexistence of one or more clusters and a subset of desynchronized elements.
If there are $K$ clusters, the fraction of elements in the system belonging to clusters is $p=\sum_{\nu=1}^K  n_\nu /N$, while 
$(1-p)N$ is the number of desynchronized elements. 

(iii) \textit{Desynchronized or incoherent}. A desynchronized or incoherent state occurs when $x^i_t \neq x^j_t$, $\forall i,j$ in the system.

\begin{figure}[h]
\includegraphics[scale=0.32]{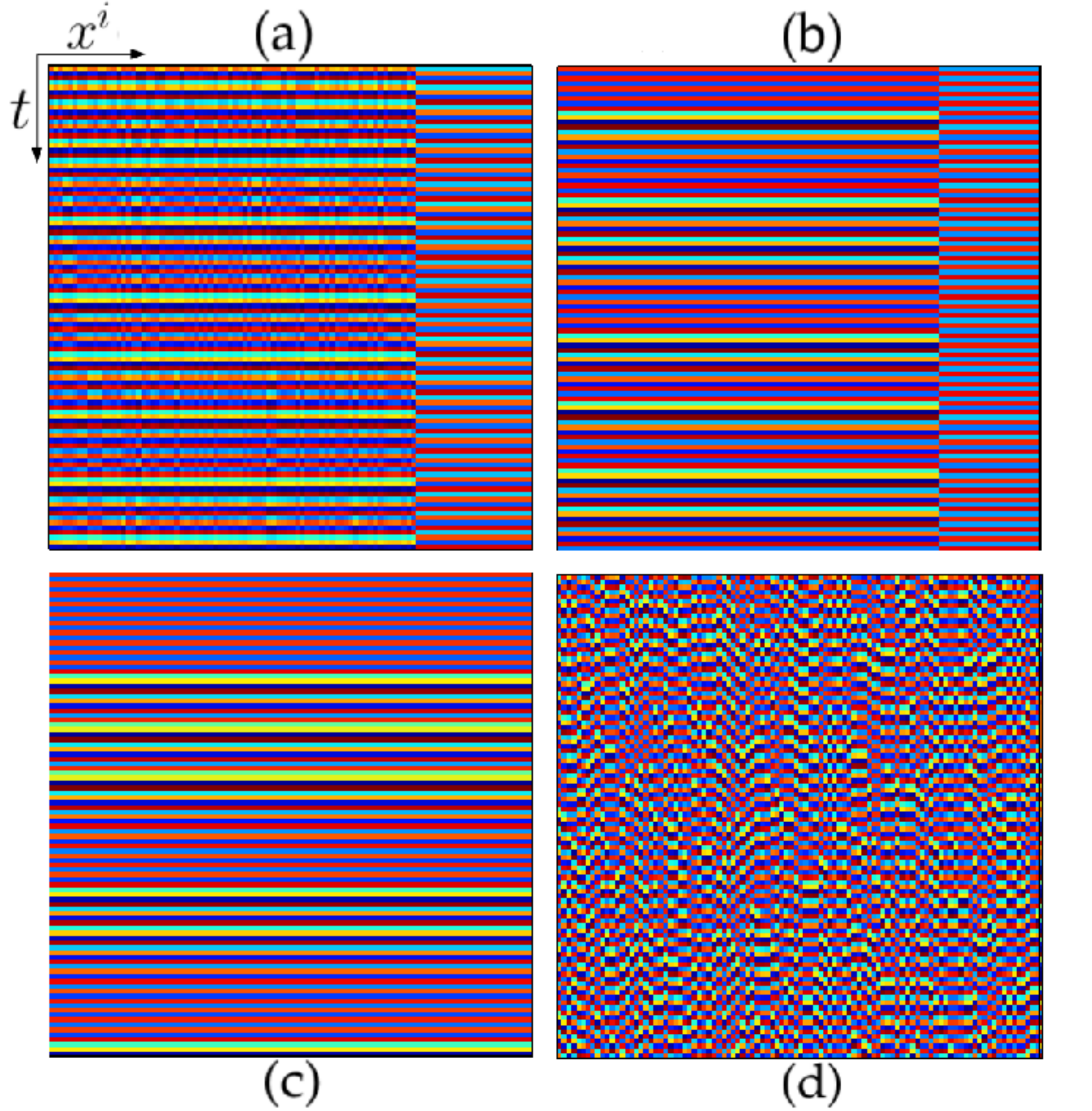}
\caption{Asymptotic evolution of the states $x^i$ (horizontal axis) as a function of time (vertical axis) for the system
of equations~(\ref{SystL1})-(\ref{SystL2})
with size $N=100$ and fixed $\alpha=1.4$ and $\beta=0.3$, for different values of the coupling parameter $\epsilon$. 
(a)  $\epsilon=0.17$, chimera state. (b)  $\epsilon= 0.21$, two-cluster chaotic state. (c)  $\epsilon=0.45$, synchronization.
(d)  $\epsilon= 0.15$, desynchronized state.
Initial conditions $x_0^i$ and $y_0^i$ are randomly and uniformly distributed in the interval $[-1,1]$  
After discarding $10^4$ transients, $100$ iterates $t$ are displayed.}
 \label{f2}
\end{figure}

Figure~\ref{f2} shows the temporal evolution of the variables $x_t^i$ of the system of equations~(\ref{SystL1})-(\ref{SystL2})  
for different values of the coupling parameter. For visualization, the indices $i$ are assigned at time $t=10^4$ 
such that $i<j$ if $x^i_t < x^j_t$ and kept fixed afterward.
The values of the states $x^i_t$ are represented by color coding. 
A chimera state and a two-cluster state are shown in Figs.~\ref{f2}(a) and \ref{f2}(b), respectively.
A chaotic synchronization state is displayed in Fig.~\ref{f1}(c), while 
a desynchronized state is shown in Fig.~\ref{f2}(d). 

In general, the number of clusters, their sizes, and their dynamical behavior (periodic, quasiperiodic or chaotic) depend 
on the initial conditions and parameters of the system.
Chimeras and clusters can be regarded as different cases of the cluster formation phenomenon: 
A cluster state consists of a few clusters $K \ll N$ of large sizes, while
a chimera state has many clusters $K=O(N)$, with a one or few
cluster of large size $n_1=O(N/2)$ and the rest of sizes  $n_\nu=1$, $\nu=2,\ldots,K$ \cite{Kaneko2}. 

In practice, we consider that a pair of elements $i$ and $j$ belong to a cluster at time $t$ if the distance 
between their state variables, defined as
\begin{equation}
 d_{ij}(t)=|x_t^i-x_t^j|,
\end{equation}
is less than a threshold value $\delta$, i.e., if $d_{ij} < \delta$. The choice of $\delta$ should be appropriate for 
achieving differentiation between closely evolving clusters. Here we use $\delta=10^{-6}$.
Then, we calculate the fraction of elements that belong to some cluster at time $t$ as
\begin{equation}
p(t)= 1-\frac{1}{N}\sum_{i=1}^N \prod_{j=1, j\neq i}^N \Theta \left( d_{ij}(t)-\delta \right),
\end{equation}
where $\Theta(x)=0$ for $x<0$ and $\Theta(x)=1$ for $x\geq 0$. We refer to $p$ as the asymptotic time average 
(after discarding a number of transients) of $p(t)$ for a given realization of initial conditions. 
Then, a clustered state in the system can be characterized by the value $p=1$.
The values $p_{\mbox{\scriptsize min}} < p < 1$ characterize a chimera state, where $p_{\mbox{\scriptsize min}}$ 
is the minimum cluster size to be taken into consideration. In this paper,
we set $p_{\mbox{\scriptsize min}}=0.05$. 

A synchronization state corresponds to the presence of a single cluster of size $N$ and it also possesses the value $p=1$. 
To distinguish a synchronization state from a cluster state, we calculate the asymptotic time average
$\sigma$ (after discarding a number of transients) of the instantaneous standard deviations of the distribution
of state variables, defined as
\begin{equation}
\sigma (t)= \left[ \frac{1}{N} \sum_{i=1}^N (x_t^i- \bar x_t)^2 \right]^{1/2},
\end{equation}
where
\begin{equation}
\bar{x}_t=\frac{1}{N} \sum_{j=1}^N x^j_t.
\end{equation}
Then, a synchronization state in the system is characterized by the values $\sigma=0$ and $p=1$, while
a cluster state corresponds to $\sigma>0$ in addition to $p=1$. A chimera state is given by
$p_{\mbox{\scriptsize min}} < p < 1$ and  $\sigma>0$. An incoherent state corresponds to $p \to 0$ and $\sigma>0$.

\begin{figure}[h]
\includegraphics[scale=0.32,angle=0]{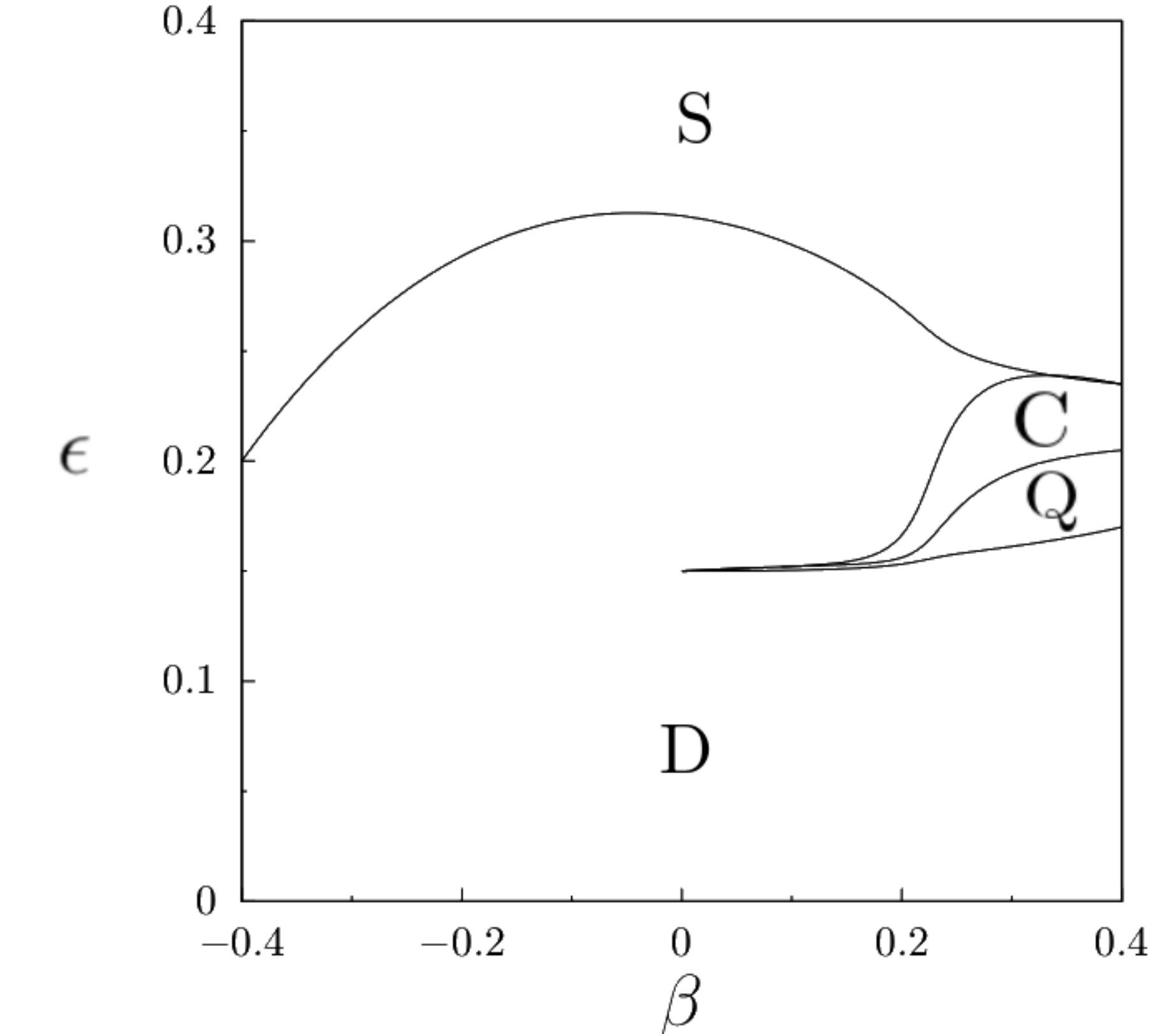}
\caption{Phase diagram on the space of parameters $(\epsilon,\beta)$ for the collective behavior of the globally coupled system 
of equations~(\ref{SystL1})-(\ref{SystL2}) with size $N=1000$ and  
fixed parameter $\alpha=1.4$. For each data point we obtain the mean values $\langle p \rangle $ and $\langle \sigma \rangle$
by averaging the asymptotic time-averages $p$ and $\sigma$
(after discarding $10^4$ transients) over $50$ realizations of initial conditions.
For each realization, initial conditions $x^i_0$ and $y^i_0$ are randomly and uniformly distributed on the interval $[-1,1]$.
Labels indicate different collective states: S: synchronization, C: cluster states, Q: chimera states, and D: desynchronization.}
\label{f3}
\end{figure}

Figure~\ref{f3} shows the collective synchronization states for 
the globally coupled system of equations~(\ref{SystL1})-(\ref{SystL2}) on the space of parameters $(\epsilon,\beta)$, characterized 
through the mean values $\langle p \rangle $ and $\langle \sigma \rangle$, obtained by averaging the asymptotic time averages 
$p$ and $\sigma$ over several realizations of initial conditions.  
Parameters $\alpha$ and $\beta$ are set in the region where robust chaos exists for the local Lozi maps. 
Synchronization occurs for large enough values of the coupling parameter $\epsilon$. 
For $\beta>0$, cluster and chimera states regions appear adjacent to each other for an intermediate range of values of $\epsilon$ . 
On the other hand, for negative values of $\beta$ only
synchronization and desynchronization can be obtained in the globally coupled system of equations~(\ref{SystL1})-(\ref{SystL2}). 
We have verified that, for values of $\alpha$ for which there is a transition FP-Chaos-P2 from the fixed point to chaos to the period-2 orbit
in Fig.~\ref{f1},
the phase diagram of the system is qualitatively the same as in Fig.~\ref{f3}.
Chimera states mediate between clusters and incoherent behavior in the parameter space.
Thus, a direct transition from complete synchronization to a chimera state is not possible in this system. 

Chimera states, referred to as partially ordered phase \cite{Kaneko2}, and cluster states were also located adjacent to each other
in the phase diagram of the globally coupled logistic map system studied by Kaneko \cite{Kaneko1}.
The local map employed in Ref. \cite{Kaneko1} did not display robust chaos, in contrast to the Lozi map used here.
The existence of periodic windows in the individual maps  
was conjectured to be a necessary condition for 
the emergence of periodic clusters 
in a globally coupled system of those maps \cite{Us2,Mikhailov}. Our results reveal that clusters, as well as chimera states,
can occur in globally coupled map networks even when the individual dynamics possesses a hyperbolic chaotic attractor or robust chaos.

At the local level, each element in the autonomous globally coupled system
of equations~(\ref{SystL1})-(\ref{SystL2}) 
is subject to the same field $h_t$ that eventually induces a collective state.
It has been shown \cite{Us1} that 
the local dynamics in a system of globally coupled maps 
can be described as a single map subject to an external signal that evolves in time identically as the field $h_t$. 
In particular, the system of equations~(\ref{SystL1})-(\ref{SystL2}) 
can be associated with a set of $N$ realizations for different initial conditions of a single 
driven Lozi map.  

\section{Nonlocally coupled Lozi maps}
In order to study the influence of the range of the interactions on the occurrence of chimera states, we consider a system
of nonlocally coupled Lozi maps described by
\begin{eqnarray}
\label{SystR1}
x_{t+1}^i&=& f(x_t^i,y_t^i) + \epsilon h_t^i   \\
y_{t+1}^i&=& \beta x_t^i, \\
h_t^i&=& \frac{1}{2k} \sum_{j=i-k}^{j=i+k} \left[ f(x_t^j,y_t^j) - f(x_t^j,y_t^j)\right],
\label{SystR2}
\end{eqnarray}
where the elements $i=1,\ldots,N$ are located on a ring with periodic boundary conditions, $\epsilon$ is the coupling parameter, 
$k$ is the number of neighbors coupled on either 
side of site $i$, and $h_t^i$ is the local field acting on element $i$. We employ the quantity $r=k/N$ to express the range of the interactions. 
Then, the value $r=0.5$ corresponds to the globally coupled system considered in Eqs.~(\ref{SystL1})-(\ref{SystL2}).

\begin{figure}[h]
\includegraphics[scale=0.32]{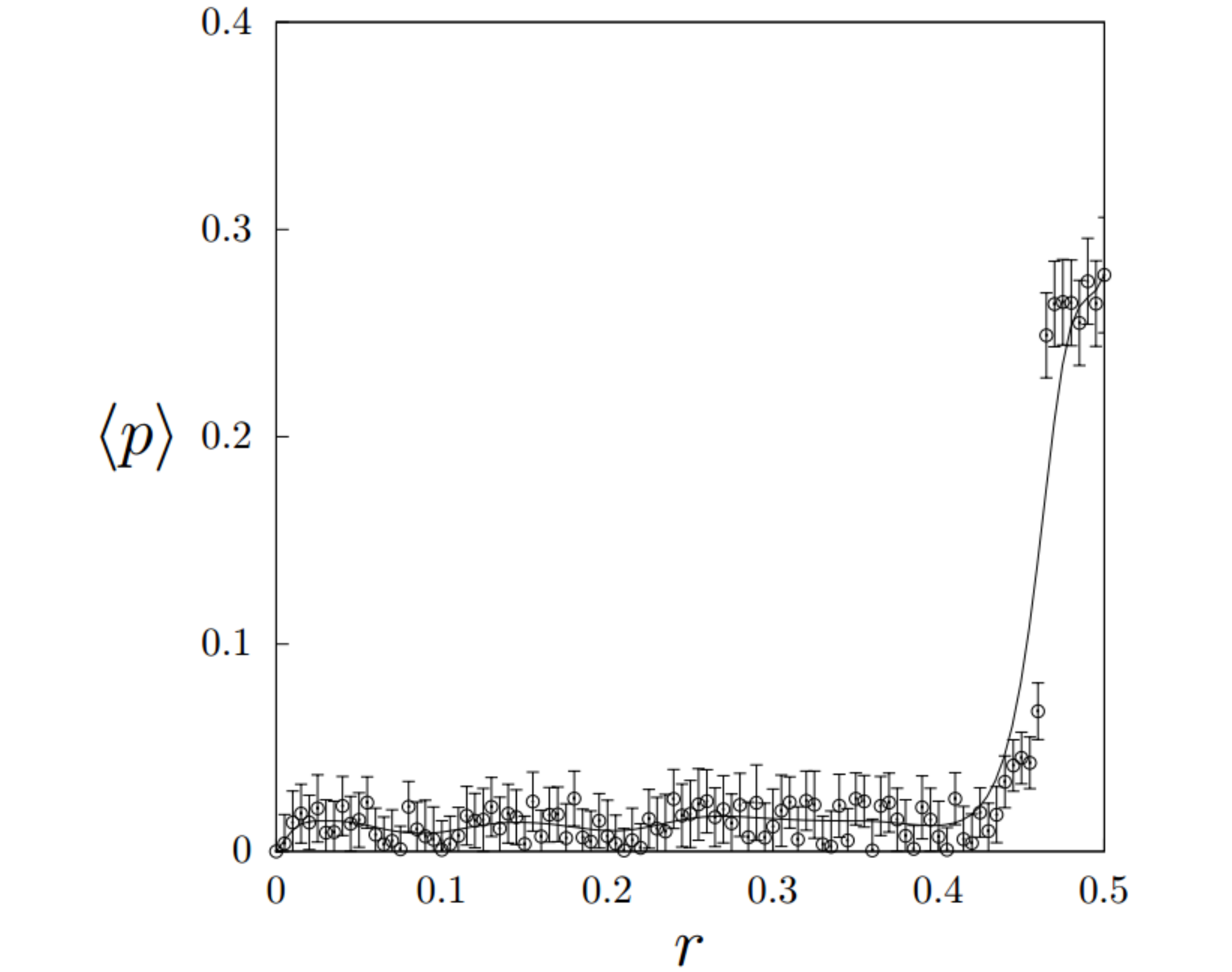}
\caption{Mean value $\langle p \rangle$ as a function of the range of interaction $r$ for the system of equations~(\ref{SystR1})-(\ref{SystR2}), 
with fixed parameters $\alpha=1.4$, $\beta=0.3$, $\epsilon=0.17$, size $N=1000$, and resolution $\Delta r=0.005$.
Each value of $\langle p \rangle$ is obtained by averaging over $100$ realizations of initial conditions, after discarding $10^4$ transients. 
Error bars indicate the corresponding standard deviations.}
\label{f4}
\end{figure}

To characterize the presence of chimera states in the system of equations~(\ref{SystR1})-(\ref{SystR2}), 
we calculate the mean value of the fraction $p$ over a number of realizations of initial conditions $(x_0^i,y_0^i)$, denoted by $\langle p\rangle$.
Figure~\ref{f4} shows $\langle p \rangle$ as a function of the range of interactions $r$ for the system of equations~(\ref{SystR1})-(\ref{SystR2}).

We observe that chimera states, corresponding to $p_{\mbox{\scriptsize min}} < \langle p \rangle < 1$,  appear for $r\geq 0.45$, that is, when $h_t^i \to h_t$.
Thus, global or sufficiently long range interactions can induce chimera states in networks of coupled chaotic hyperbolic maps.
We have verified that cluster states can also be achieved for large enough values of the range $r$ in 
the system of equations~(\ref{SystR1})-(\ref{SystR2}). 

\section{Conclusions}
The presence of chimera states in globally coupled networks of identical oscillators seemed at first counterintuitive because of the 
perfect symmetry of such a system \cite{Pik}. 
However, such networks are among the simplest extended systems that can exhibit chimera behavior. 
We have shown that the presence of global interactions can indeed allow for the emergence of chimera states
in networks of coupled elements possessing chaotic hyperbolic attractors, such as Lozi maps, where such states do not form with local interactions. 
We have employed two statistical measures to characterize different collective states of synchronization
in the space of parameters of the globally coupled system: chimera states, cluster states, complete synchronization, and incoherence. 
With an appropriate ordering of the indexes of the maps, we were able to visualize the spatiotemporal patterns corresponding to these states. 
Additionally, we have shown that chimera states can appear in arrays of nonlocally coupled Lozi maps with a sufficiently long range of interactions.

We have found that chimeras are closely related to cluster states in this system of globally coupled Lozi maps, a feature 
that has been observed in other globally coupled systems \cite{Schmidt,Kaneko1}. Since dynamical cluster formation is typical in
many systems with global interactions, one may expect that the phenomenon of chimera states should also be commonly found in such systems,
including those possessing other hyperbolic chaotic attractors.
Our results suggest that chimera states, like other collective behaviors,
arise from the interplay between the local dynamics and the network topology; either ingredient can prevent or induce 
its occurrence.

\section*{Acknowledgments} 
This work was supported  by Project No. C-1906-14-05-B of Consejo de Desarrollo Cient\'ifico, Human\'istico, Tecnol\'ogico
y de las Artes, Universidad de Los Andes, M\'erida, Venezuela. M. G. C. is grateful to the Associates Program
of the Abdus Salam International Centre for Theoretical Physics, Trieste, Italy, for visiting opportunities.

\end{document}